\documentclass[fleqn,usenatbib]{mnras}

\usepackage[T1]{fontenc}
\usepackage{ae,aecompl}
\usepackage{xcolor}

\usepackage{graphicx}	
\usepackage{amsmath}	
\usepackage{amssymb}	

\title[The impact of the L factor on the Fermi paradox]{
A probabilistic analysis of the Fermi paradox in terms of the Drake formula: the role of the L factor}

\author[N. Prantzos]{
N. Prantzos \thanks{E-mail: prantzos@iap.fr}
\\
Institut d'Astrophysique de Paris and Sorbonne Universit\'e,  98bis Bd Arago, 75014 Paris, France 
}

\date{Accepted XXX. Received YYY; in original form ZZZ}

\pubyear{2015}

\begin{document}
\label{firstpage}
\pagerange{\pageref{firstpage}--\pageref{lastpage}}
\maketitle

\begin{abstract}
In  evaluating the number of technological civilizations N in the Galaxy through the Drake formula,  emphasis is mostly put on the astrophysical and biotechnological factors describing the emergence of a civilization and much less on its the lifetime, which is intimately related to its demise. 
It is argued here that this factor is in fact the most important regarding the practical implications of the Drake formula, because it determines the maximal extent of the "sphere of influence" of any technological civilization. The Fermi paradox is studied in the terms of a simplified version of the Drake formula, through Monte Carlo simulations of N civilizations expanding  in the Galaxy during their space faring lifetime L. In the framework of that scheme, the probability of "direct contact" is determined as the fraction of the Galactic volume occupied collectively by the "spheres of influence" of N civilizations. The results of the analysis are used to determine  regions in the parameter space where the Fermi paradox holds.  It is argued that in a large region of the diagram the corresponding parameters suggest rather a "weak" Fermi paradox.  Future research may reveal whether a "strong" paradox holds in some part of the parameter space.
Finally, it is argued that the value of N is not bound by N=1 from below, contrary to what is usually assumed, but it may have a statistical interpretation.
\end{abstract}




\section{ Introduction}
\label{sec:Intro}

Most of the effort in establishing the chances for the existence of life elsewhere in the universe has been put on evaluating the various astronomical and biological factors that may lead to the emergence of life on the surface of terrestrial-type exoplanets around other stars in the Milky Way. Twenty  five years after the discovery of the first exoplanet, we have now started to have sufficient data to allow us to think that such objects are relatively common in the Galaxy \citep{Petigura2018}. However, we have not yet any evidence about the existence of other lifeforms, even at the microscopic level, elsewhere in the Universe.

Metrodorus, disciple of Epicurus, seems to have been the first to formulate the main argument for the existence of life forms and intelligent beings beyond Earth,  in the third century BC: "to consider the Earth as the only populated world in infinite space is as absurd as to assert that in an entire field sown with millet only one grain will grow". The idea of an infinite space, populated by an infinite number of atoms and their combinations, was a key ingredient of the atomistic philosophy of Leucipus, Democritus, and Epicurus \citep{Furley1987}.

Today, proponents of extraterrestrial intelligence (ETI) invoke essentially the same argument, although the concept of infinity is not used any more (because it is difficult to handle and it may lead to paradoxes, e.g., in an infinite universe, everything, including ourselves, could exist in an infinite number of copies). But the number of stars in our Galaxy, $\sim$100 billion, is considered by some  - mostly astronomers - to be large enough as to make Metrodorus' argument applicable to the Milky Way. Others, however - evolutionary biologists, in particular - are not impressed by that number and remain skeptical concerning ETI
\citep[e.g.][]{Simpson1964,Mayr1992}.

In the past sixty years or so, the debate on ETI was largely shaped by the "Drake equation" and the "Fermi paradox". The former -  which should rather be called "Drake formula" - was proposed  by Frank Drake \citep{Drake1961} and became ever since the key quantitative tool to evaluate the probabilities for radio-communication with extraterrestrial intelligence (CETI). The latter - also known as "Fermi's question" - was formulated a decade earlier by  Enrico Fermi, but it remained virtually unknown until 1975, when it was independently re-discovered twice \citep{Hart1975,Viewing1975}. In a concise form ("{\it where are they?"}) it opposes a healthy skepticism to the optimistic views on ETI:  if there are many of them, why don't we see any evidence of their presence on Earth or in its neighborhood?
\cite[see ][for a recent comprehensive
overview of the subject ]{Cirkovic2018}.

It is not known what kind of calculations - if any - Fermi  did  to evaluate the chances of the Earth being visited by extraterrestrial civilizations, during a lunch-time discussion with colleagues at Los Alamos in 1950 \citep[see][for an account of that discussion]{Jones1985}. Several attempts have been made to quantify Fermi's question through numerical simulations, based on various assumptions \citep{Newman1981,Jones1981,Fogg1987,Landis1998,Bjork2007,Hair2013,Zackrisson2015,Carroll2019}. These studies reached different conclusions on the paradoxicality of Fermi's question, both from the quantitative and the qualitative point of view, e.g. the timescales required for colonization of the Galaxy \citep{Webb2015,Ashworth2014}; however, none of those works  makes explicit use of the Drake formula.

In this work, a new framework is introduced for the study of Fermi's question, using a simplified form of Drake's formula (Sec. \ref{sec:Drake}) to evaluate the number N of technological civilizations. In this framework, the possibility of N$<$1 is explored and is given a statistical interpretation (Sec. \ref{sec:Lonely}). It
 is argued that, in the steady state situation expressed by Drake's formula,  the lifetime $L$ plays a key role for the study of Fermi's question, because the volume of the "sphere of influence" of communicating civilizations (either through direct contact or through ELM signals) depends on L$^3$ for an isotropic expansion. This dependence is illustrated by Monte Carlo simulations performed  for various values of the relevant parameters N, L and $\upsilon$ (the typical  speed of the expansion front of the civilizations), in a disk galaxy with the dimensions of the Milky Way (Sec. \ref{sec:Fermi}). The main result is that the  probability of contact can be defined as the fraction of the Galactic volume occupied by the common volume of the spheres of influence of the ensemble of galactic civilisations.   This original presentation allows one to display in terms of the Drake formula and in a compact form, the three physical explanations mostly discussed as solutions to the Fermi paradox: civilizations are rare, too short-lived, or unable to expand at sufficiently high speed in the Galaxy (Sec. \ref{sec:fermi_criterion}).

\section{A fresh look at Drake's formula}
\label{sec:Drake}

On the occasion of a now famous meeting -the first one on the Search for Extraterrestrial Intelligence (SETI) - that he organized in Green Bank, Virginia, Drake tried to evaluate the number of radio-communicating civilizations in our Galaxy \citep{Drake1961}.  In its original formulation, the Drake equation reads
\begin{equation}
N \ = \ R_* \ f_p \ n_e \ f_l \ f_i \ f_t \ L 
\label{eq:Drake_ori}
\end{equation}
where R* is the rate of star formation in the Galaxy (i.e., number of stars formed per unit time), $f_p$ is the fraction of stars with planetary systems, $n_e$ is the average number of planets around each star, $f_l$ is the fraction of planets where life developed, $f_i$ is the fraction of planets where intelligent life developed, and $f_T$ is the fraction of planets with technological civilizations. Obviously, N and L are intimately connected: if N is the number of radio-communicating civilizations - as in the original formulation by Drake - then L is the average duration of the radio-communication phase of such civilizations (and not their total lifetime, as sometimes incorrectly stated). On the other hand, if N is meant to be the number of technological or space-faring civilizations, then L represents the duration of the corresponding phase \citep[see][for additional comments on the Drake formula ]{Prantzos2013}.

The Drake formula obviously corresponds to the equilibrium solution of an equation similar to the  equation of radioactivity for the decay rate D of a number N of radioactive nuclei: D = $dN/dt$ = - N/L, where L is the lifetime of those nuclei. In the steady state, where the production rate P is equal to the decay rate D, one has N = P L. In a similar vein, the product of all the terms of the Drake formula, except $L$, can be interpreted as the production rate P of technological (radio-communicating or   space-faring) civilizations in the Galaxy\footnote{The same formula is obtained within a different framework, the Little's law \citep{Little1961}, well known in probability theory and statistics.}.

On the basis of this analogy, it was  suggested \citep{Prantzos2013} that the practical implications of the Drake formula would be more clearly evaluated if its original seven terms were condensed to only three. The aim was twofold: (1) to illustrate quantitatively some implications of the number N for SETI and CETI, and (2) to use exactly the same framework for a quantitative assessment of the Fermi paradox. 
The Drake equation is now written as
\begin{equation}
N \ = \ R_{\rm {ASTRO}} \ f_{\rm BIOTEC} \ L    
\label{eq:Drake_modi}
\end{equation}
where
\begin{equation}
 R_{\rm {ASTRO}} \ = \ R_* \ f_p \ n_e
 \label{eq:Drake_Rastro}
\end{equation}
represents the production rate of habitable planets, and 
\begin{equation}
 f_{\rm BIOTEC} \ =    \ f_l \ f_i \ f_t \
 \label{eq:Drake_Fbiotec}
\end{equation}
represents the product of all chemical, biological and sociological factors leading to the development of a technological civilization.

Obviously, while the product  $R_{\rm {ASTRO}}  f_{\rm BIOTEC}   $ represents the "production term" of technological civilizations, the lifetime $L$ represents the "destruction term" in the steady state situation described by the Drake formula.

From the three terms of the modified Drake equation (Eq.
\ref{eq:Drake_modi}) $R_{\rm {ASTRO}}$
is the only one reasonably well studied at present and expected to be well constrained in the foreseeable future.  
The first of its terms, R$_*$, is already constrained by observations in the Milky Way to be ~4 stars/yr. The present day star formation (SF) rate  is ~1.9 M$_{\odot}$/yr \citep{Chomiuk2011} and there are ~2 stars per M$_{\odot}$ in a normal stellar initial mass function (IMF) like the one of \cite{Kroupa2002}. However, its average past value was probably higher by a factor of 2, and we shall adopt the value of 4 M$_{\odot}$/yr which corresponds to an average star production rate of  <R>$\sim$8 stars yr$^{-1}$; this average SF rate reproduces well the stellar mass of ~4 10$^{10}$ M$_{\odot}$  or the $\sim$10$^{11}$ stars of the Milky Way if assumed to hold for the age of the Galaxy A$\sim$10 Gy. 

We shall assume that only 10\% of those stars are appropriate for harboring habitable planets, because their mass has to be smaller than 1.1 M$_{\odot}$, i.e. they have to be sufficiently long-lived (with main sequence lifetimes larger than $\sim$4 Gy as to provide enough time for the development of intelligence and technology) and larger than 0.7 M$_{\odot}$, to possess circumstellar habitable zones outside the tidally locked region  \citep{Selsi2007}. This leaves aside the most numerous class of stars, namely the low mass red dwarfs: their intense and time-varying activity \citep[e.g.][and references therein]{Paudel2019} and the limitations on their circumstellar habitable zone \citep{Haqq2018,Scwhieterman2019} largely balance the effect of their larger number with respect to solar-type stars. In any case, considering them would increase the planet numbers by factor of 10-20 but this would change little the conclusions, given the extremely large uncertainties of the other factors of the Drake equation, as illustrated in a recent analysis  \citep{Wandel2019}.

A recent analysis of the Kepler DR25 and Gaia DR2 data \citep{Hsu2019} finds that the statistics currently available on extra-solar planets around solar-type (FGK) stars point to ~20\% of the surveyed stars possessing planets with sizes 1-1.75 Earth radii and orbital periods of 237-500 days.  This fraction may be considered to describe the product $f_p n_e$ in the Drake equation. It is, admittedly, a rather optimistic estimate, its only merit being that it imposes a plausible upper limit on the fraction of such solar-type stars. Combined with the aforementioned formation rate of 0.7-1.1 M$_{\odot}$ stars, it leads to $R_{\rm ASTRO}\sim$0.1 habitable planet per year, which is adopted in this study.
Even if the value is revised by a factor of a few (most probably downwards) by future assessments, the results presented here could be easily scaled accordingly.

The term $f_{\rm BIOTEC}$  is not constrained from below. It has a maximum  value of $f_{\rm BIOTEC,Max}$ = 1 (corresponding to  $ \ f_l = f_i = f_t \ $ =1), which is rather implausibly optimistic but constitutes a useful upper limit.  Because of that, the number of N technological civilizations at steady state is bound from above by the value  N$_{\rm MAX}$=$R_{\rm ASTRO} L$ = 0.1 $L$ (with $L$ expressed in yr) in the Drake formula.

\begin{figure}
	\includegraphics[width=0.47\textwidth]{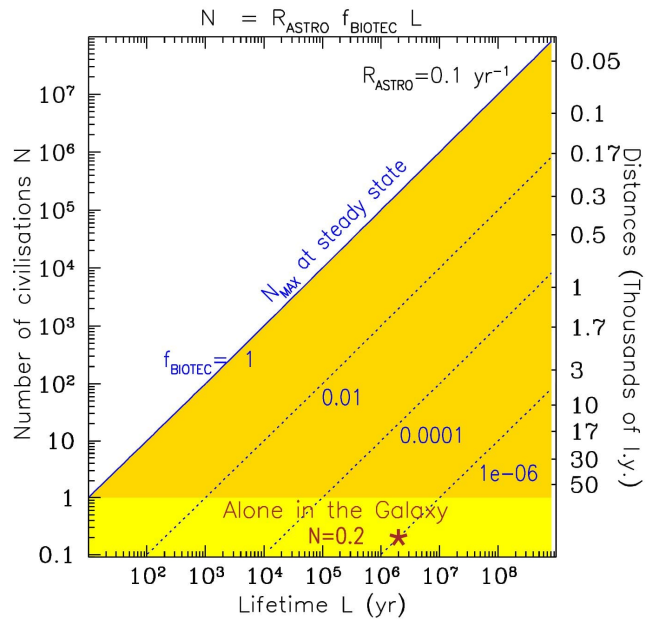}
    \caption{ Number of civilizations N versus their average lifetime L, assuming $R_{\rm ASTRO}$ = 0.1/yr.
    Straight lines correspond to biotechnological factors $f_{\rm BIOTEC}$=1 (solid), 10$^{-2}$, 10$^{-4}$ and 10$^{-6}$ (dotted).  Typical distances D (in light-years l.y.) between civilizations for each N are obtained from the discussion of Sec. \ref{sec:Drake} and are indicated on the right axis. The bottom (yellow shaded) region corresponds to N $<$ 1. The statistical significance of that region (illustrated by the point N = 0.2) is discussed in Sec. \ref{sec:Lonely}.  }
    \label{fig:NvsLandDist}
\end{figure}

The re-formulated Drake equation (Eq. \ref{eq:Drake_modi}) appears in a graphical form, in Fig. \ref{fig:NvsLandDist}, after fixing $R_{\rm ASTRO}$=0.1 y$^{-1}$ and plotting N  as a function of $L$. A slightly different form of this diagram ($f_{\rm BIOTEC}$ vs $L$) appeared in \cite{Prantzos2013}, but this one is more straightforward and illustrates the situation in a clearer way. In this log-log diagram, values of  N  vs $L$ for a given $f_{\rm BIOTEC}$ are represented by  straight quasi-diagonal lines. Taking into account that  $f_{\rm BIOTEC}\leq$1, the solid line at  $f_{\rm BIOTEC}$=1 bounds N from above, i.e. there are no civilizations on the upper left part of the diagram in the steady state situation. Same values of N are obtained for different combinations of  $f_{\rm BIOTEC}$ and $L$, but we argue that the dependence on L is more important regarding its implications.

On the right axis of the figure appear, on a different scale, the typical distances  between such civilizations. Indeed, as noted in \cite{Prantzos2013},
the Drake formula was  meant to evaluate the chances of establishing radio-communication with ETI in the Galaxy, but in fact it says very little on the probability of contacting them: number N alone  is totally insufficient to evaluate that probability.  From the point of view of interstellar communication or direct contact  it is a completely different  matter to have 10 civilizations inside the whole Galaxy or inside a globular cluster. For that purpose, one more step has to be taken, to  connect the lifetime $L$ to the typical distances between such civilizations, in order to account for the finite speed of electromagnetic (ELM) signals (or transport); those distances depend on N and on the dimensions of the system, in this case the Milky Way. 

To evaluate typical distances between Galactic civilizations it may be assumed that, to a first approximation, the Galactic disk is described by a cylinder of radius R$_{\rm G}$ =12 kpc ($\sim$40 000 l.y.) and height h$_{\rm G}$=1 kpc ($\sim$3 260 l.y.), where the N civilizations of the Drake formula are distributed uniformly\footnote{ A better approximation would be to consider the exponential profile of the stellar disk of the Milky Way, but the conclusions depend little on such assumptions (factors of order unity) and much more on the unknown factors of the Drake formula.}. By equating the volume of the Galactic cylinder V$_{\rm G}$=$\pi$ R$_{\rm G}^2$h$_{\rm G}$ with the sum of N volumes of spheres of average radius $r$ occupied by each civilization , one obtains the average distance between two civilizations as 

\begin{equation}
 D \ = \ 2 \ r \ = \ 2  \ \left ( \frac{3 \rm h_G R_{\rm G}^2}{4 N} \right )^{1/3}  \ \ \ (D<{\rm h_G})
\end {equation}
In the case of a small number of civilizations (say N<1000) it turns out that D>h$_{\rm G}$ and the appropriate expression is then 
\begin{equation}
D \ = \ 2 \ r = \ 2 \ {\rm R_{ G}}/\sqrt(N)  \ \ \ (D>{\rm h_G})
\label{eq:distances}
\end {equation}

It is interesting to notice that even for a hundred thousand civilizations, typical distances are of several hundred l.y., making contact - either by radio-signals or space probes  - difficult.

Notice that in the simplified picture presented here, it is  assumed that all civilizations have similar values of $L$, that is, the dispersion $\Delta L$  is $\Delta L \ll L$. This need not be the case. Statistical treatments, considering $\Delta L \sim L$,  and canonical distributions have been studied \citep{Maccone2010,Glade2012}. However, in any case, the unknown mean value of $L$ plays a more important role than the equally unknown form of its distribution. 

\section{Lonely hearts in the Galaxy?}
\label{sec:Lonely}

\begin{figure}
	\includegraphics[width=0.499\textwidth]{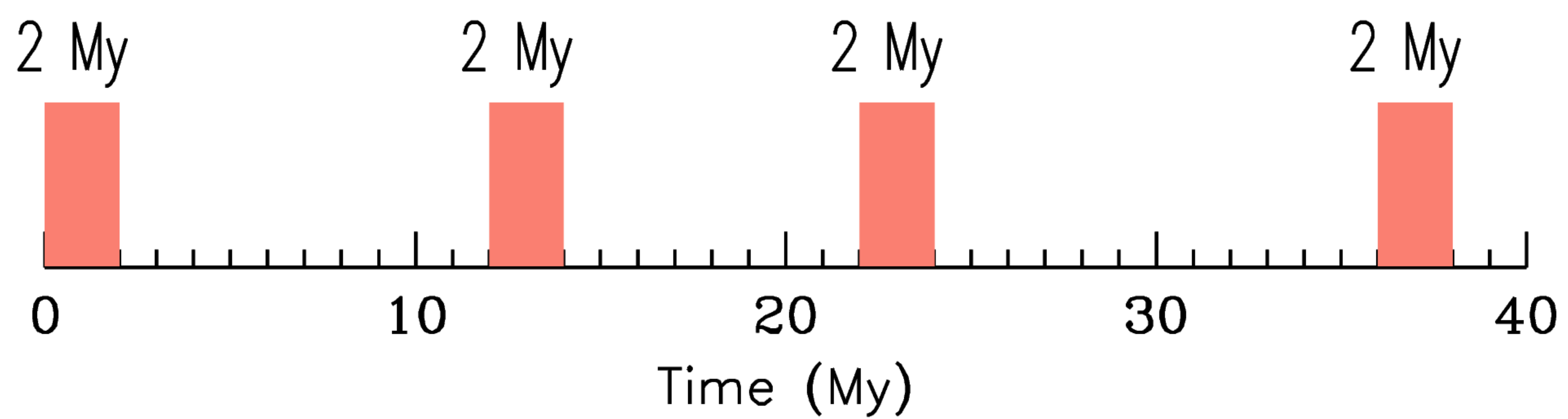}
    \caption{ Statistical interpretation of the case N=0.2, illustrated in the bottom panel of Fig. \ref{fig:NvsLandDist}. L/N is the typical timescale of appearance of civilisations of lifetime L in the Galaxy.
     }
    \label{fig:Lonely}
\end{figure}

The region below the line N = 1 is not necessarily void. It corresponds to values of N < 1 which are considered impossible, since our own civilisation exists. However, such values may have a physical - albeit probabilistic - meaning, as illustrated in Figure \ref{fig:Lonely}: they may represent the fraction f of the time span T between the appearance of two successive civilizations that is occupied by the typical lifetime $L$: f = $L$/T. In Fig. $\ref{fig:Lonely}$, this is illustrated by assuming arbitrarily that $L$ = 2 My and T = 10 My. Four civilizations appear within 40 million years, and last for 2 My each. Their summed lifetime is 8 My, that is, they exist for 8/40 = 0.2 of that time span. For an external observer, the probability of finding a technological civilization anytime in the Galaxy is then 0.2 and this number may be considered as the {\it number N} of civilizations of lifetime $L$ in the Galaxy at a given moment assuming steady state.

The case N<1 is rather depressing. A technological civilization may emerge in the Galaxy and live thousands or millions of years, being alone during its whole existence and separated from both its predecessors and its successors by huge time gaps. It is possible that thousands of technological civilizations, much more advanced than ours, blossomed in the Galaxy and reigned for thousands or millions of years. During that time, they  may have explored  a large fraction of their neighborhood, seeking  for civilisations still alive or  gone long ago, through artefacts and other traces. And they ultimately disappeared, unable to find others, unaware that others have preceded them and forever unknown to those who succeeded them - especially if they did not venture far away from their solar systems. Each one of them was found to be "alone in space" - at least in the space covered by the Milky Way - but not "alone in time", since others may have preceded and succeeded them.
Our own civilization may be such a "lonely heart" in the Milky Way: not "the first", not "the only one", but simply alone.

The implications of that situation would be obvious for any sufficiently advanced civilization: only a systematic research by unmanned probes or manned spaceships could provide in the long run evidence for or against the existence of other  lifeforms \citep[see e.g.][and references therein, for a recent overview]{Crawford2018}. Indeed, many technological civilizations may be unwilling or unable to communicate with others or to explore their outer space, or they may simply be extinct by now, in a still life-bearing planet.

\begin{figure*}
	\includegraphics[width=0.9\textwidth]{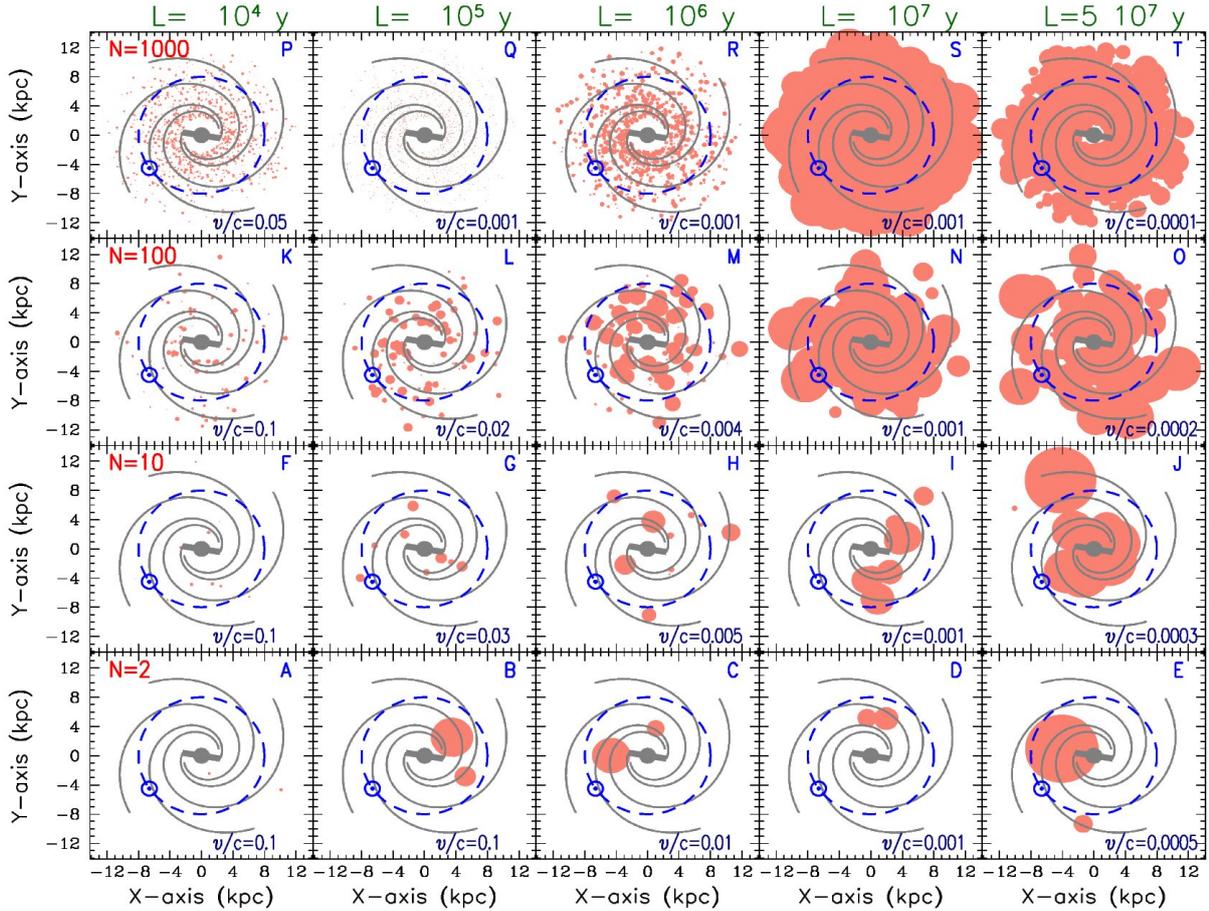}
    \caption{ Monte Carlo simulations of N civilizations of lifetime $L$ in steady state, appearing in the Galaxy at random places and times and expanding spherically during time t<L at speed $\upsilon$/c (where c is the speed of light), for various values of N (=2, 10, 100, and 1000 from bottom to top) and $L$ (=10$^4$, 10$^5$, 10$^6$, 10$^7$ and 5 10$^7$ y, from left to right). 
    The values of  $\upsilon$/c are given in the bottom left of each panel. The Sun is located by the blue symbol at 8 kpc from the Galactic center and the solar circle is indicated by the blue dashed curve. The four-arm spiral pattern and the bar of the Galaxy are (approximately) indicated by grey curves. The size of the filled circles indicates the projected aerea of the "sphere of influence" of each civilization  covered during its age t<$L$ at speed $\upsilon$/c.}
    \label{fig:MC_sim}
\end{figure*}

\section{Expansion in the Galaxy and the Fermi paradox}
\label{sec:Fermi}

 A simple method to evaluate the conditions under which the Fermi paradox holds in the framework of the Drake formula was suggested in  \cite{Prantzos2013}: the paradox is assumed to hold when the N civilizations, expanding in the Galaxy  at a fraction  $\beta$ of the speed of light c, are able to cover collectively within their "spheres of influence" the whole volume of the Galaxy during their lifetime $L$.

 The adopted scheme for galaxy colonization is known as the "coral model" and was suggested in 2007 by J. Benett and S. Shostak \citep[latest edition]{Bennett2017} and further developed (including a statistical description) in \cite{Maccone2010}.  It assumes that once a civilization masters the techniques of interstellar travel, it starts a thorough colonization/exploration of its neighborhood for its whole lifetime $L$. Colonization proceeds in a directed way, i.e. it concerns only stars harboring nearby habitable planets, which are detected before the launching of the spaceships. Ships are sent to new stars not from the mother planet but from the colonized planets in the colonization front and they are launched after some time interval Dt following colony foundation. This gives enough time to the colonizers to install on the new planet and prepare the next colonizing mission. Notice that  $\upsilon$ is the effective velocity of the colonization front and not the velocity of the interstellar ships, which has to be larger than  $\upsilon$ (to compensate for the time required for installation and preparation of new missions, but also for the fact that the new missions would not always head "outwards" but also "sideways" for a full exploration of the neighborhood). Notice also that the lifetime $L$ corresponds to a single civilization and includes all its offspring colonies; in other terms, the colonies do not count as different civilizations, since they do not originate in an independent way.
 
Fig. \ref{fig:MC_sim} provides illustrations of this scheme, through several Monte Carlo simulations for various combinations of the values of the involved parameters.  It is assumed that N civilizations emerge at random places in the Milky way disk\footnote{A simulation along those lines was recently performed in \cite{Grimaldi2018} to explore probabilities of SETI through radio-signal research.} and at random times during the last period of duration $L$. In other terms, at any time  a flat  distribution of ages is obtained. The spatial distribution  of N takes into account the stellar surface density profile of the Galactic disk ($\propto$ exp(-R/3 kpc)), i.e. their surface density is larger in the inner disk. After their emergence as a technological species, civilizations are assumed to start expanding in the Galaxy filling progressively around them spheres of radius $r=\upsilon t$ and volume $V=4/3 \pi r^3$, where $t$ is their current age (0<$t$<L); for $r>{\rm h_G}$, the volume is   $V= \pi {\rm h_G} r^2$, i.e. once the height of the galactic disk is reached, civilizations expand  radially in the plane of the disk and no more vertically to it.

The parameters N=2, 10,100 and 1000 are adopted from the bottom to the top of Fig. \ref{fig:MC_sim}, while L =10$^4$, 10$^5$, 10$^6$,  10$^7$ and 5 10$^7$ y are used  from left to right. The value of expansion speed  $\upsilon$ (in units of light speed c) appear in each panel. It is emphasized that this value is the one of the expansion of the "colonization front" and not the one of the spaceships, which should be necessarily higher. For that reason, the value   $\upsilon$/c=0.1 is rather an optimistic upper limit and is used here only for illustration purposes. In the various schemes studied so far for Galactic colonisation, speeds $\upsilon$/c between 0.1 and 10$^{-5}$ have been considered (see Table 1 in \cite{Fogg1987} for an early compilation) and we use here values within that range. Smaller values would obviously require larger values for N and/or $L$ in order to compensate and result in colonization of the Galaxy within $ L$.

The six panels with the same value of $\upsilon$/c=0.001, i.e. panel D with (N, $L$)= (2, 10$^7$), panel I (10, 10$^7$), panel N (100, 10$^7$), panel S (1000, 10$^7$), panel R (1000, 10$^6$) and panel Q (1000, 10$^5$), illustrate the importance of factor L: increasing number N by a factor of 100 (from 10 to 1000, going from I to S) while keeping $L$ the same- that is, by increasing the factor $f_{\rm BIOTEC}$ by 100 - obviously increases proportionally the fraction of the Galactic volume\footnote{It should be emphasized that the term "explored galactic volume" means "explored stars within that volume".}  covered collectively by the N civilizations by that same factor. But an increase of L by a factor of 100 for the same N (i.e. going from panel Q to S) increases that fraction by a much larger factor ( $\sim$10$^4$). This is the reason why from the various factors of the Drake equation, $L$ is the most important numerically. For sufficiently large $L$, around  10$^7$ y, even a "small" number of civilisations (N$\sim$100) expanding at "modest"  speeds - of the order of 1 l.y. per thousand years - can collectively occupy a large fraction of the Galaxy.

This scheme reproduces also naturally situations that have been already described in completely different contexts in order  to explain the Fermi paradox. 
\cite{Ostriker1986} argued that, even if advanced technological civilizations are common, they are unlikely to fully occupy the Galaxy, because at some point of their expansion, their mutual interactions could reduce the pace of colonization, leaving some portions of the Galaxy unoccupied for periods of the order of $L$. They base their arguments on a mathematical analysis drawing from the ideas of theoretical ecology and they suggest that the Earth may be found in such an unoccupied region, thus providing  another explanation of the Fermi paradox.
\cite{Landis1998} performed simulations based on percolation theory to simulate the expansion of civilizations throughout the Galaxy. Depending on the adopted parameters of his model, he found that large unoccupied regions may be found within colonized volumes, thus explaining the Fermi paradox. Recently, \cite{Carroll2019} adopted a multi-parameter scheme - taking also into account the natural motion of stars within the Galaxy - to show that clusters of continuously occupied systems, as well as quasi-void regions could co-exist in the Milky way. 

All the aforementioned works and many others \citep[see e.g. the Introduction in][] {Carroll2019} adopt several extraneous parameters, beyond the factors of the Drake formula, and those parameters play a crucial role in the final outcome of the corresponding studies. We think that for the first time  a framework using explicitly and exclusively the parameters of the Drake formula is adopted; the addition of one more parameter, the speed of the expansion front, is mandatory in order to place the discussion in the context of the Milky Way taking into account its size.

\section{A quantitative criterion for Fermi's paradox}
\label{sec:fermi_criterion}

Our scheme provides a quantitative answer to Fermi’s question in terms of probabilities: assuming that there are at present N civilizations in the Galaxy, appearing randomly in space and time  and exploring their neighborhood for time $L$ at speed  $\upsilon$,  the probability that our solar system is "currently" (i.e. in the last $L$) within at least one of their  "volumes of influence" $V_i$ can be  defined as the fraction $F$ of the Galactic volume V$_{\rm G}$ occupied by the ensemble of those volumes:
\begin{equation}
 F(N,L,\upsilon) \ = \ \frac{\sum_{i=1}^{N} V_i}{{\rm V_G}}
\end{equation}
For large N, the sum tends to N$<V>$, where the average volume is 
\begin{equation}
\begin{split}
  <V>  & =  \frac{1}{L} \ \int_0^L V(L) dL   \\
       & =  \frac{1}{L} \frac{4 \pi}{3 }  \int_0^L (\upsilon t)^3 dt = \frac{\pi}{3} (\upsilon L)^3  \ \ \ \ \ \ {\rm for} \ \ (\upsilon L<{\rm h_G})   \\
       & = \frac{1}{L} \pi \ {\rm h_G}  \int_0^L (\upsilon t)^2 dt = \frac{\pi \rm h_G}{3} (\upsilon L)^2  \ {\rm for} \ \ (\upsilon L>{\rm h_G})
  \end{split}
  \label{eq:Av_vol}
\end{equation}
This leads to
\begin{equation}
\begin{split}
  F(N,L,\upsilon) & = \ \frac{ N \ <V>}{{\rm V_G}} \\
 & = \ \frac{1}{3} \ \frac{ N \ (\upsilon \ L)^3}{ {\rm h_G \ R_G^2}}  \ \ \ \ {\rm for} \ \ (\upsilon L<{\rm h_G})  \\
  & = \ \frac{1}{3} \ \frac{N \ (\upsilon \ L)^2}{ {\rm R_G}^2} \ \ \ \ {\rm for} \  (\upsilon L>{\rm h_G})
  \end{split}
  \label{eq:Prob}
\end{equation}



In fact, the dependence of the probability on $L$ is even stronger than it appears from Eq. \ref{eq:Prob}, because N is not an independent variable, it depends on $L$ according to Eq. \ref{eq:Drake_modi} where the independent factor is $f_{\rm BIOTEC}$ rather than N. Replacing N in  Eq. \ref{eq:Prob} through its expression of  Eq. \ref{eq:Drake_modi} leads to a dependence of the probability on  $L^3$ or $L^4$: 

\begin{equation}
\begin{split}
  F(f_{\rm BIOTEC},L,\upsilon) 
 & = \ \frac{1}{3} \ \frac{ R_{\rm ASTRO} \ f_{\rm BIOTEC} \ L \ (\upsilon \ L)^3}{ {\rm h_G \ R_G^2}}  \ (\upsilon L<{\rm h_G})  \\
  & = \ \frac{1}{3} \frac{ R_{\rm ASTRO} \ f_{\rm BIOTEC} \ L  \ (\upsilon \ L)^2}{ {\rm R_G}^2} \  (\upsilon L>{\rm h_G})
  \end{split}
  \label{eq:Prob2}
\end{equation}

\begin{figure}
	\includegraphics[width=0.47\textwidth]{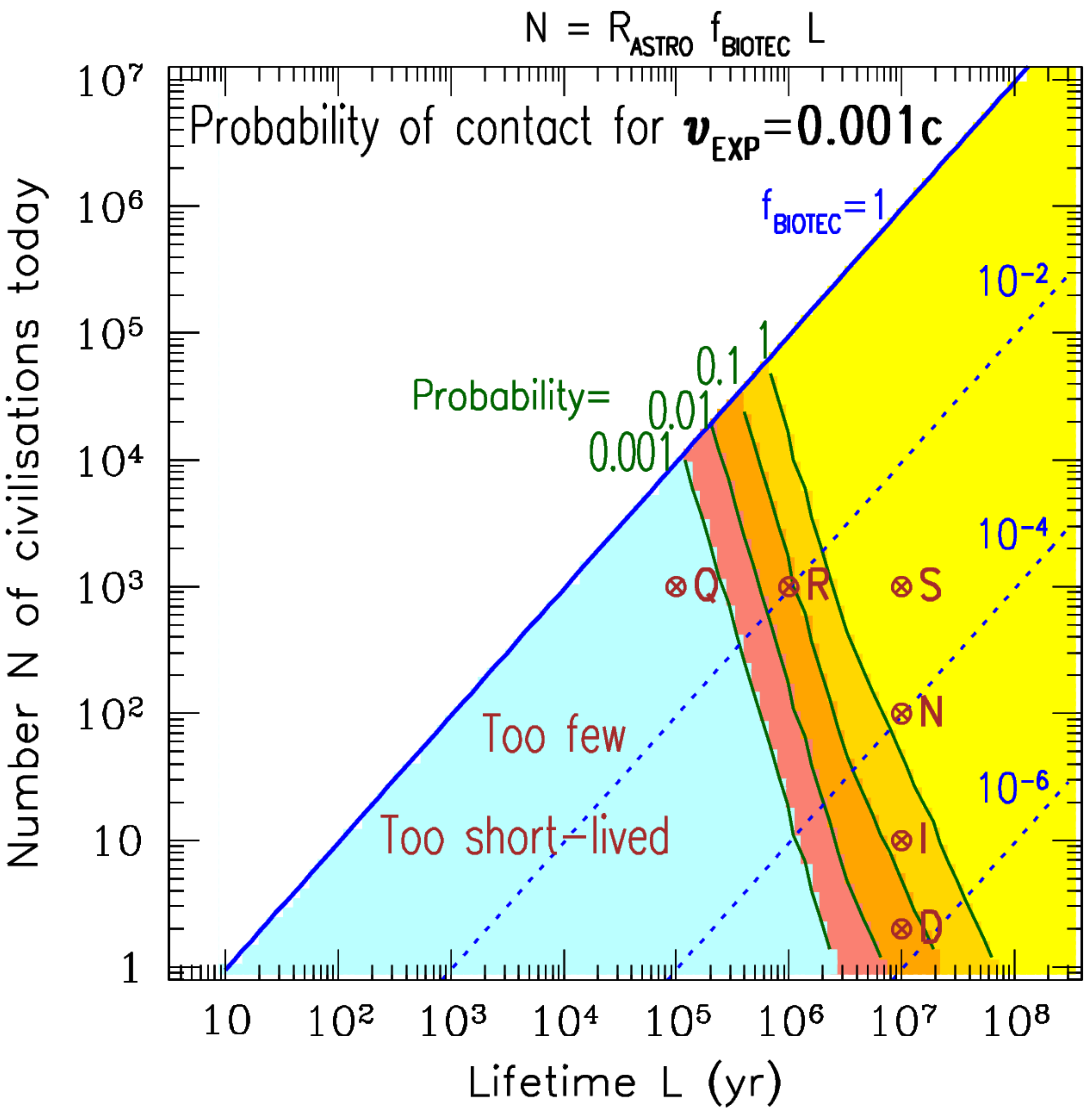}
    \caption{ Results of a systematic investigation of the N vs $L$ plane for $\upsilon$/c=0.001, assuming a formation rate of habitable planets $R_{\rm ASTRO}$ =0.1 yr$^{-1}$ in the Milky Way. Blue diagonal lines indicate the number N as function of $L$ and correspond to $f_{\rm BIOTEC}$ =1 (maximum value, solid line),  10$^{-2}$, 10$^{-4}$ and 10$^{-6}$ (dotted blue lines). The probability of the N civilizations covering the  Galactic volume  with their «spheres of influence»  is colour coded. Probabilities of  10$^{-3}$, 10$^{-2}$, 10$^{-1}$ and 1 are indicated by the green thin curves. In the yellow region to the right of the figure, the probability is $\geq$1. To its left, the probability of contact decreases with $L^{-3}$, to become negligible in the cyan shaded region.
     The six models of Fig. \ref{fig:MC_sim} which adopt $\upsilon$/c=0.001 appear with the corresponding letters (see the text).
  }
    \label{fig:Probability}
\end{figure}

Thus, although  $f_{\rm BIOTEC}$  and $L$ enter the Drake formula linearly, the latter has a much greater impact on the Fermi paradox than the former: a variation  in $f_{\rm BIOTEC}$ by four orders of magnitude can be compensated by a variation of $L$ by a factor of ten. Despite that, we shall keep our subsequent  discussion in terms of the (N,$L$) variables, which are more intuitive.

An illustration of our quantitative evaluation for the case $\upsilon$/c=0.001 is provided in Fig. \ref{fig:Probability} for $R_{\rm ASTRO}$ =0.1 y$^{-1}$.   The solid line at $f_{\rm BIOTEC}$ =1 bounds N from above, i.e. there are no civilizations on the upper left part of the diagram in the steady state situation. The yellow shaded region to the right corresponds to $F \geq$1, indicating full coverage of the Galaxy and even overlapping spheres of influence\footnote{No attempt is made here to evaluate the impact of overlapping spheres of influence on the outcome of the calculation, i.e. would the contact between civilizations stop the expansion of one of them (or both) or, on the contrary, would it accelerate that expansion?} . This region indicates the parameter space whereas Fermi’s question does not admit a physical answer: {\it they should be here but we don't see them, so where are they ?} However, the values required for the various parameters are rather high. If space-faring technology is trivially developed in the Galaxy ($f_{\rm BIOTEC}\sim$1),  10$^5$ civilizations should co-exist (upper right part) each one expanding for more a million years  at one thousandth of the speed of light in order to fully cover the Galactic volume and thus render the observed absence of contact with them truly problematic. The situation would be equally problematic in the case of rare technology ($f_{\rm BIOTEC}\sim$10$^{-6}$), requiring a dozen civilizations to expand at the same speed but for about 10$^8$ years (lower right part). 

Regions to the left of the yellow region correspond to progressively lower probabilities, as indicated by the three solid green curves at $F$=1, 10$^{-1}$, 10$^{-2}$ and 10$^{-3}$, where civilizations are too few or too short-lived to collectively colonize the whole Galaxy. The importance of the factor $L$ is clearly seen:  a reduction of L by a factor of 10 decreases the probability of covering the Galactic volume by a factor of 100-1000, making contact rapidly improbable. On the other hand, a decrease of $f_{\rm BIOTEC}$ by a factor of 100 can be easily compensated by an increase of $L$ by a factor of 5-7, as can be seen in the case of each of the green  probability curves. In the cyan shaded region to the left of the curve  $F$=10$^{-3}$ the probability is practically zero. Fermi's question can be easily understood in physical terms in that region, without invoking any "sociological" reasons (e.g. unwillingness to explore or contact us, cosmic "zoo" etc.)

\begin{figure}
	\includegraphics[width=0.47\textwidth]{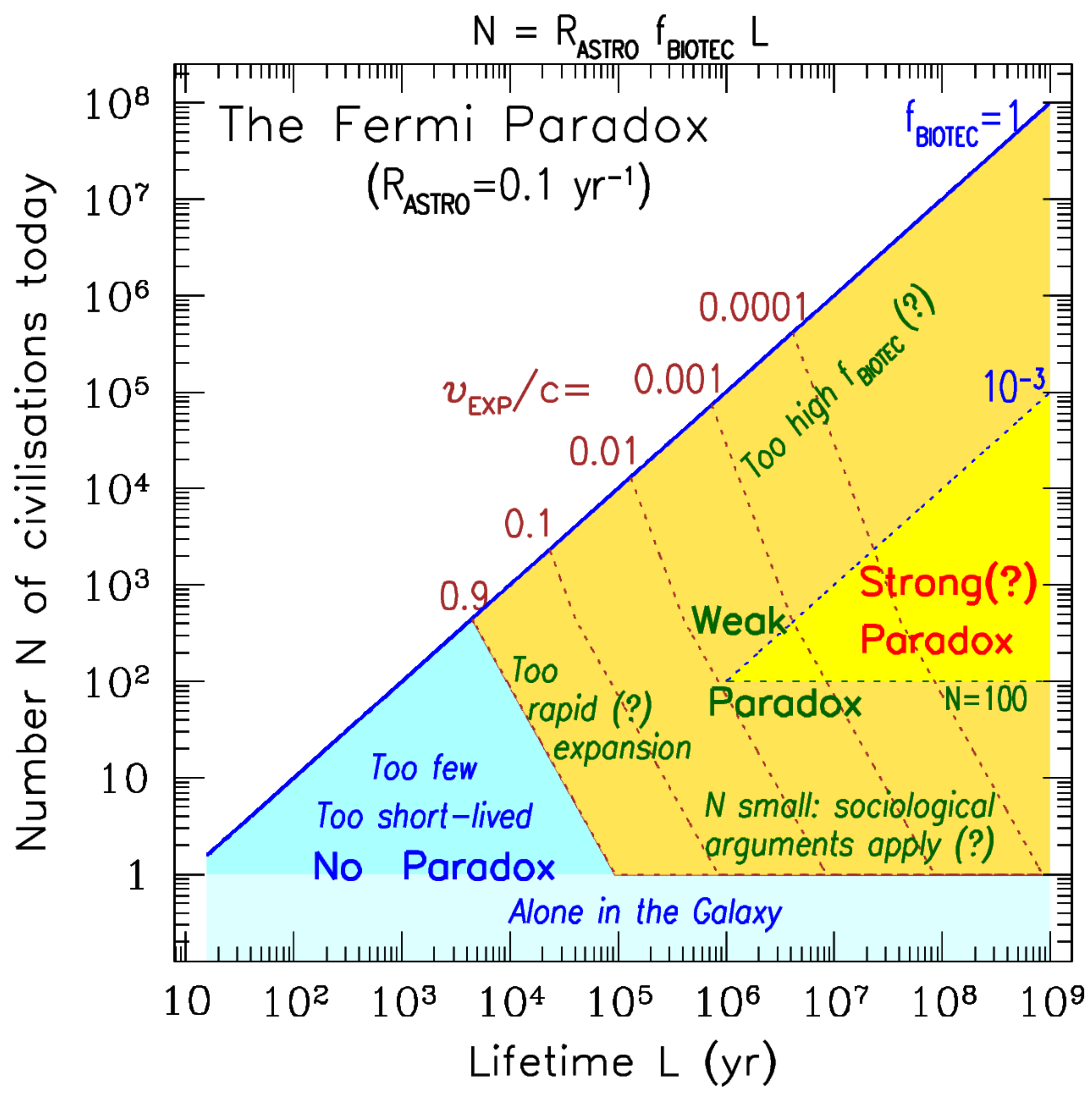}
    \caption{ The Fermi paradox presented in the N vs. $L$ plane, in terms of the Drake formula and  for $R_{\rm ASTRO}$ =0.1 yr$^{-1}$ in the Milky Way.  It is assumed that the colonization front expands with average speed $\upsilon$/c=0.9, 0.1, 0.01, 0.001 and 0.0001, as indicated by the dotted curves.  The Fermi paradox holds to the right part of each speed curve, i.e. the probability of a collective colonization of the Galaxy by N civilizations is =1 in that region. In the region in the bottom left part  the Fermi question can be understood in physical terms, since the probability of collective occupation of the Galaxy is practically zero ("No paradox"). To its right, probabilities are formally high, but conditions are rather too "optimal" in some cases (thus a "Weak paradox"). It remains to be seen whether there are regions of the parameter space where conditions are "reasonable", allowing one to qualify Fermi's question as a "Strong  paradox" (see the text).} 
    \label{fig:Fermi}
\end{figure}

Each of the lettered points  in Fig. \ref{fig:Probability} indicate the corresponding models in Fig.  \ref{fig:MC_sim}, which adopt $\upsilon$/c=0.001. Again, as discussed before,  the role of $L$ is clearly and quantitatively illustrated: the probability $P$ decreases by a factor of 10$^4$ when $L$ decreases by 100 (going from Model S to Q, for the same N=1000), but it decreases more slowly, by a factor 100 when N decreases also by 100 (going from S to D, for the same $L$=10$^7$).

Simulations for different values of $\upsilon$/c, from 0.9 to 0.0001, are summarized in Fig. \ref{fig:Fermi}: the shaded area to the right of each curve, identified by the corresponding $\upsilon$/c, is the region of $F \geq$1 where the criterion for having a "Fermi paradox" is satisfied. It is essentially a simple  "geometrical" criterion which can be expressed as

\begin{equation}
\begin{split}
   & N \ (\upsilon \ L)^3 \geq  \ 3 \ {\rm h_G \ R_G^2}  \ \ \ \ {\rm for} \ \ (\upsilon L<{\rm h_G})  \\
  & N \ (\upsilon \ L)^2  \geq \ 3 \ {\rm R_G}^2 \ \ \ \  \ \ \ {\rm for} \  (\upsilon L>{\rm h_G})
  \end{split}
  \label{eq:Criterion}
\end{equation}
for the case of the "coral model" adopted here. To the left of each curve, the probability of contact rapidly decreases (the inequalities of Eq. \ref{eq:Criterion} are inversed) and there is no Fermi paradox for the corresponding value of  $\upsilon$/c (as in Fig. \ref{fig:Probability}). The case of $\upsilon$/c=0.9 is shown only for illustration purposes, as it is improbable that civilisations may expand at such high speed (and even at $\upsilon$/c=0.1), because the spaceships should travel even faster than that.

In \cite{Prantzos2013} it is argued that if N is "small" (arbitrarily taken to be N<100) the dozens of various "sociological" hypothesis put forward to explain Fermi’s paradox \citep{Webb2015} may really constitute an answer for the absence of extraterrestrials on Earth. In that case, one may consider that Fermi's question does not constitute a true paradox. Also, it may be that expansion of the "exploration front" at speeds $\upsilon$/c>0.01 is too rapid for any technology, since the actual speed of the spaceships should be even larger. 

Furthermore, the "biotechnological" factor $f_{\rm BIOTEC}$ must certainly be  lower than its maximal possible value of 1. For illustration purposes, we have put the value  $f_{\rm BIOTEC}$=10$^{-3}$ in Fig. \ref{fig:Fermi}, but the actual value may turn out to be much smaller. 
Future  observations may reveal the presence of life elsewhere \citep{Kopparapu2019}  probing thus the value of $f_l$, the first of the three factors of the term  $f_{\rm BIOTEC}$ in Eq. \ref{eq:Drake_Fbiotec}. On the other hand, evolutionary biologists argue for values of the intelligence factor $f_i$ much lower than 1 \citep[see][and references therein]{Lineweaver2009}, in which case the region of the "strong paradox" would shrink much more.

For all those reasons, it is argued that in a fairly large region of the diagram (orange shadowed in Fig. \ref{fig:Fermi}) Fermi's question constitutes only a "weak paradox". It remains to be seen whether there are  regions to the right of the diagram where Fermi's question can be considered as really meaningful ("strong paradox"), i.e. : {\it if there are so many, live so long and expand at not an unreasonable speed, why aren’t they here?} 

From Fig. \ref{fig:Fermi} it appears that, for "realistic" expansion speeds ($\upsilon$/c$<$0.01) and independently of their number, civilizations should live more than $L \sim$1 My in order to fully cover the Galaxy in their lifetime.   Background extinction rates of species on Earth suggest lifespans of the order of 1 My for mammals or 11 My for invertebrates \citep{Lawton1995}, shrinking even further the "strong paradox" region from the right part in Fig. $\ref{fig:Fermi}$. However, it is unknown at present whether such considerations could apply to extraterrestrial and/or technologically advanced species.

In any case, the adopted scheme allows one to asses Fermi's question in the phase-space of  quantifiable - albeit unknown - parameters ($f_{\textsc {BIOTEC}}$, $L$ and $\upsilon$), accounting also for the astrophysical setting i.e. the size of the Galaxy. But this "quantification" obviously cannot help with other important issues related to the Fermi paradox, e.g. "{\it to assert  the presence of extraterrestrials on Earth in the recent or distant past, what kind of tracers should we seek for?}"\citep[see e.g.][and references therein]{Schmidt2019}. The lack of a convincing answer to that question  obviously limits the utility of Fig. \ref{fig:Fermi}.

\section{Summary}
\label{sec:summary}

In this work the Fermi paradox is analyzed in terms of a simplified version of the Drake formula - originally suggested in \cite{Prantzos2013} - and the role of the civilization lifetime $L$ is emphasized. Several novelties are introduced in the discussion.

a) The condensed form of Drake's formula (N = R$_{\rm ASTRO} f_{\rm BIOTEC} L$) is presented graphically in the plane N vs $L$,  assuming that the "astronomical" factor R$_{\rm ASTRO}$ can be determined from present-day and forthcoming observations; based on current understanding, $R_{\rm ASTRO}$=0.1 yr$^{-1}$ is adopted throughout this work. The plane is covered by the different values assumed for the "biotechnological" factor $f_{\rm BIOTEC}$ and N is bound from above by the value N=0.1 $L$ (where $L$ is expressed in yr), since  $f_{\rm BIOTEC,Max}$=1.

b) The possibility of N$<$1 is explored and for the first time a statistical interpretation is suggested: it corresponds to the case where the typical lifetime $L$ of a civilization is smaller than the typical timescale T of the emergence of two successive civilizations in the Galaxy, i.e. N=$L$/T. With this interpretation, the Drake formula covers the case of civilizations being "alone in space" (within the Milky Way), but not "alone in time".

c) It is argued that, in the steady state situation expressed by Drake's formula,  the lifetime L plays a key role, even larger than $f_{\rm BIOTEC}$, despite the fact that both factors enter the formula in a linear way. The reason is that  the volume of the "sphere of influence" of communicating civilizations (either through direct contact or through ELM signals) depends on $L^3$  for an isotropic expansion, at least in the framework of the "coral model" for Galactic colonization adopted here.

d) This dependence is illustrated by Monte Carlo simulations performed  for various values of the parameters N, $L$ and $\upsilon$ (the typical  speed of the expansion front of the civilizations), in a disk galaxy with the dimensions of the Milky Way.

e) A quantitative criterion is proposed to evaluate the chances of contact: the probability of contact is the fraction of the Galactic volume occupied by N "volumes of influence"  during the last period of duration $L$.  Eq. \ref{eq:Prob} and \ref{eq:Prob2}  reveal the strong influence of the factor $L$ on the discussion of the Fermi paradox. 

f) This criterion allows one to define in the plane N vs $L$ regions where the probability of contact is high or low, for a given assumed value of the expansion speed $\upsilon$ (Fig. \ref{fig:Probability}). For sufficiently large values of N and $L$, a probability is P$\geq$1 may be  obtained, indicating that Fermi's question calls for answer: {\it if they are so numerous and expand for so long and sufficiently rapidly, then why are they not here?} The case P$>$1 implies overlapping of some volumes of influence. No attempt is made here to interpret the implications of such overlapping, i.e. of the contact between civilizations. 

g) On the basis of the above, a criterion quantifying the Fermi paradox is proposed (Eq. \ref{eq:Criterion}) relating  the two variables of the Drake formula ($N$ and $L$) to the size of the Galaxy through the expansion speed $\upsilon$.

h) This original presentation allows one to display quantitatively in a single figure (Fig. \ref{fig:Fermi}) and in a compact form, the three physical explanations mostly discussed as solutions to the Fermi paradox\citep[see][and references therein]{Webb2015}: rare civilizations (low $N$), too short-lived (low $L$, making them unable to arrive here, even if they are numerous and/or expand rapidly), or unable to expand at sufficiently  high speed (low $\upsilon$). 

 i) It is argued that in a large region of the diagram the corresponding parameters suggest rather a "weak" Fermi paradox. Future research may reveal whether a "strong"paradox holds in some part of the parameter space. In any case, 
 it appears that for "realistic" expansion speeds ($\upsilon$/c$<$0.01)  civilizations should expand for more than $\sim$1 My in order to fully cover the Galaxy in their lifetime.

It should be emphasized that the quantitative results obtained here depend on three key assumptions:

a) {\it The number N of Galactic civilisations is obtained by  the Drake formula and reflects a steady-state.}
In principle, a different framework may be conceived, in which the "production rate" of civilisations is not connected to the birth rate of habitable planets or does not correspond to a steady state: for instance, they may appear in "waves" in time and/or in space following some "special" event, e.g. the passage of a spiral wave or some other perturbation. Such considerations would make the analysis considerably - and unnecessarily, at this stage - more complex.

 b) {\it Civilisations appear randomly in the Galactic volume } (the MC simulations of Fig. \ref{fig:MC_sim} take into account the radial stellar density profile, but the analytical criteria of Eq. \ref{eq:Criterion} do not); the situation would be obviously different if, for some reason, some Galactic places are systematically favoured. 

c) {\it The expansion front  expands as in the "coral model" } and its radius increases as $r_{\rm exp} \propto L$, leading to a strong dependence of the "volume of influence" on $L$ ($V_{\rm exp} \propto L^3$). In other models, like the diffusion model adopted in e.g. \cite{Newman1981}, the radius increases as $r_{\rm exp} \propto \sqrt{D L}$ - where $D$ is the diffusion coefficient - and the dependence on $L$ is weaker ($V_{\rm exp} \propto L^{3/2}$). Although it is trivial to calculate such models in the adopted framework (replacing the appropriate quantities in Eq. \ref{eq:Av_vol} to \ref{eq:Criterion}), it is hard - and certainly not intuitive - to decide about the values of the diffusion coefficient $D$ and even about the physical meaning of such models. 

For several decades, the Drake formula played an important role in the search for extraterrestrial life, providing a framework to formulate our current understanding about a very complex phenomenon such as the development of life and intelligence in the astrophysical setting of the Milky Way
\citep{Prantzos2000}.  In this study, we show that it can also be used to  constrain quantitatively the "physical" answers to Fermi's question. Forthcoming  developments in various fields related to astrobiology, space sciences, communication theory, big data analysis etc. are expected to enrich further our understanding of this topic \citep[e.g.][and references therein]{Cabrol2016}.

\section*{Acknowledgements}

I am grateful to my colleague Gary Mamon at IAP for invaluable help with the graphics of this work (and many others).




\bibliographystyle{mnras}
\bibliography{Prantzos_Reference} 





\bsp	
\label{lastpage}
\end{document}